\documentclass[twocolumn,english,prd,superscriptaddress,aps,showkeys,amsmath,amssymb,floatfix,nofootinbib]{revtex4-1}
\pdfoutput=1
\usepackage[pdftex]{graphicx}
\usepackage[pdftex]{color}
\usepackage[colorlinks,bookmarks]{hyperref}
\definecolor{linkblue}{rgb}{0,0,0.8}
\definecolor{linkgreen}{rgb}{0,0.5,0}
\definecolor{darkgreen}{rgb}{0,0.4,0}
\definecolor{purple}{rgb}{0.7,0.0,0.4}
\hypersetup{linkcolor=linkblue, citecolor=purple, urlcolor=linkblue}
\usepackage{float}
\usepackage{nicefrac}
\usepackage{xcolor}

\usepackage[T1]{fontenc}
\usepackage[utf8]{inputenc}
\usepackage{lmodern}
\usepackage{babel}
\usepackage{bm}
\usepackage{verbatim}
\usepackage[us]{datetime}
\usepackage{enumerate}
\usepackage{amsmath}

\usepackage{latexsym}
\usepackage{mathrsfs}

\definecolor{vale}{rgb}{0,0.5, 1.}

\def\dd{{\rm d}}
\def\d{{\rm d}}
\def\be{\begin{equation}}
\def\ee{\end{equation}}
\def\bea{\begin{eqnarray}}
\def\eea{\end{eqnarray}}
\def\nn{\nonumber}
\usepackage{amsmath}
\usepackage{hyperref}
\usepackage{physics}
\usepackage{bm}
\begin{document}

%\title{Comparing Dark Energy models with Hubble versus Growth Rate data}

%\author{Bryan Sagredo, Javier Silva Lafaurie and Domenico Sapone}
%\affiliation{Cosmology and Theoretical Astrophysics group,\\
%	Departamento de F\'isica, FCFM, Universidad de Chile,\\
%	Blanco Encalada 2008, Santiago, Chile}

\title{Comparing Dark Energy models with Hubble versus Growth Rate data}

\author{Bryan Sagredo}
\email{bryan.sagredo@ing.uchile.cl}
%\affiliation{}

\author{Javier Silva Lafaurie}
\email{javier.silva@ug.uchile.cl}
%\affiliation

\author{Domenico Sapone}
\email{dsapone@ing.uchile.cl}
\affiliation{Cosmology and Theoretical Astrophysics group, Departamento de F\'isica, FCFM, Universidad de Chile, Blanco Encalada 2008, Santiago, Chile}

\begin{abstract}
In this work we perform an analysis on the recently proposed conjoined cosmic growth and cosmic expansion diagram~\cite{Linder:2016xer} to compare several dark energy models using the Figure of Merit showed in~\cite{Basilakos:2017rgc}, which consists in the inverse of the $1\sigma$ confidence region in the $f\sigma_8(z)-H(z)$ plot.
Our analysis also consists of comparing the models by performing different statistical criteria:  Bayes factor~\cite{Trotta:2005ar}, the Bayesian Information Criteria~\cite{Schwarz:1978tpv} and the Akaike Information Criterion~\cite{AIC2}.  We also developed a 3-dimensional Figure of Merit to account simultaneously for the errors on the growth rate and the Hubble parameter.  The main idea is to consider several cosmological models and compare them with the different statistical criteria in order to highlight the differences and the accuracies of each single criterion.

\end{abstract}

\keywords{cosmology, dark energy, model comparison}

\maketitle

\section{Introduction}

Recent observations~\cite{Riess:1998cb,Perlmutter:1998np} pointed out that the Universe seems to be in a phase of accelerated expansion. These evidences have led cosmologists to revise the theory of the expansion of the Universe either by introducing a new component called dark energy~\cite{Sapone:2010iz} or by modifying directly the theory of gravity~\cite{Tsujikawa:2010zza}.

Within the framework of Friedmann-Lema$\text{\^i}$tre-Robertson-Walker (FLRW) cosmologies, such accelerated expansion can be generated by adding up a simple cosmological constant $\Lambda$ to the total budget of the Universe. Even though the latter gives rise to severe coincidence and fine-tuning problems, observations still confirm such an explanation~\cite{Betoule:2014frx, Ade:2015xua, Abbott:2017wau}. 
Over the years a series of dark energy models have been considered in order to solve, or at least alleviate, the theoretical problems related to dark energy. However, none of these explanations seem to be convincing.  

Alternative theories of gravity came naturally as a consequence of the incapability of having a self-consistent model of dark energy. This class of models intends to modify General Relativity (GR) and to explain the observed acceleration of the Universe as a pure weakening of gravity at very large scales. 

The important question here is whether the two scenarios can be distinguished. It is well known that any Hubble expansion can be generated by choosing an appropriate equation of state for the dark energy, see~\cite{Bonvin:2006en}. However, over the years there have been claims that it is possible to distinguish alternative theories of gravity from dark energy models by using growth data; the last assumption is not always true unless the expansion history is fixed, \cite{Kunz:2006ca}. Nonetheless,  recent works have proposed to study the cosmic growth versus cosmic expansion history conjoined diagram, the $f\sigma_8-H$ plot, to put constraints on the parameter space of cosmological models, or to compare different models directly~\cite{Linder:2016xer}. Model comparison using this approach has already been investigated in~\cite{Moresco:2017hwt, Basilakos:2017rgc}. The advantage of the $f\sigma_8-H$ plot over other probes relies on the degeneracy break of the history curves when comparing different models or the parameter space, since it contrasts a geometrical observable, given by $H(z)$, to a  pure gravitational effect, given by $f\sigma_8(z)$. 

Using this approach, dark energy models were compared using the $f\sigma_8-H$ plot~\cite{Basilakos:2017rgc} through the FoM defined as the inverse area of the $1\sigma$ confidence region in the conjoined diagram. In this work, we follow a similar approach and we also compare the models using different statistical tools: the standard Bayesian evidence~\cite{Trotta:2005ar}, the Bayesian Information Criteria~\cite{Schwarz:1978tpv} (BIC), the Akaike Information Criterion \cite{AIC} (AIC) and the FoM. Furthermore, we considered an extension to the FoM which we define 3-FoM, which considers both errors on $f\sigma_8(z)$ and $H(z)$. 

Anticipating the results, we find that the FoM is a fairly good estimator of the errors, however, its extension, the 3-FoM, captures simultaneously the growth of matter and the expansion history making it more stable over different models. The criteria BIC and AIC$_c$ penalize substantially models with extra parameters. 

The paper is structured as follows: in Section \ref{basic-eqs} we report the basic equations that will be used in our work, whereas in Section \ref{models} we list the cosmological models that will be compared, and the link between $H$ and $f\sigma_8$ measurements. In Section \ref{data} we show the datasets used in the analysis and the statistical methodology is reported in Section \ref{methodology}. In Section \ref{results} we report the results of our analysis.

\section{Basic equations}\label{basic-eqs}
The evolution of a general fluid can be expressed in terms of its present energy density parameter $\Omega_{0}$ and its equation of state parameter (EoS) $w(a) = p(a)/\rho(a)$, where $p$ and $\rho$ are the pressure and energy density of the fluid, respectively, and $a$ is the scale factor, normalized to 1 today. 
The EoS is the key quantity that fully characterizes the fluid at the background level. 
%The subscript $0$ will denote the present-day value of the corresponding quantity. 

Using a general formalism the Hubble parameter in a non-flat cosmology is given by
\begin{equation}
\label{Hubble}
H^2=H_0^2 \qty[\Omega_{m_0}a^{-3}+\Omega_{k_0} a^{-2}+\Omega_{de_0}a^{-3(1+\hat w)}]\,,
\end{equation}
where $H_0$ is the Hubble constant, $\Omega_{m_0}$, $\Omega_{k_0}$, $\Omega_{de_0}$ are the present-day values of  matter, curvature and dark energy densities, respectively. 
Furthermore, the parameters satisfy the relation $\Omega_{m_0}+\Omega_{k_0}+\Omega_{de_0}=1$. The total matter density is:
\bea
\label{matter_density}
\Omega_m(a)=\left(1+\frac{1-\Omega_{m_0}}{\Omega_{m_0}}a^{-3\hat w}\right)^{-1}\,.
\eea
The quantity $\hat w$ in Eqs.~\eqref{Hubble} and~\eqref{matter_density} is the effective EoS parameter accounting for the time dependence, given by
\begin{equation}
\hat w(a) = \frac{1}{\ln{a}} \int\limits_1^a \frac{w(x)}{x}\dd{x} \,.\nn
\end{equation}
The angular diameter distance is defined as
\begin{equation}
    d_A(z)= \frac{c\,H_0^{-1}}{(1+z)\sqrt{-\Omega_{k_0}}}
    \sin( \sqrt{-\Omega_{k_0}} \int\limits_0^{z} \frac{H_0}{H(y)}\dd y)\,,\nn
\end{equation}
which reduces to
\begin{equation}
    d_A(z)=\frac{c}{H_0} \frac{1}{1+z}\int\limits_0^{z} \frac{H_0}{H(y)}\dd y\;,\nn
\end{equation}
if the curvature is set to zero.

By gravitational collapse, matter forms structures in the universe, which are called perturbations $\delta\rho(a,\,k)$, where $k$ represents the scale in Fourier space. These perturbations grow over time according to the characteristics of the fluid: EoS, pressure perturbation $\delta p$ and anisotropic stress $\sigma$. 

The growth of perturbations for a general fluid is governed, assuming homogeneity and isotropy, by the differential equations \cite{Ma:1995ey} 
\bea
\delta'=&3&\left(1+w\right)\phi'-\frac{V}{H\,a^2}-\frac{3}{a}\left(\frac{\delta p}{\rho}-w\delta\right)\,,   \label{eqs:perts-fluid-1} \\
V'=&-&\left(1-3w\right)\frac{V}{a}+\frac{k^2}{H\,a^2}\frac{\delta p}{\rho} + (1+w)\frac{k^2}{H\,a^2}\psi \nn\\
&-&(1+w)\frac{k^2}{H\,a^2}\sigma\;,
  \label{eqs:perts-fluid-2}
\eea
where the primes denote derivatives with respect to the scale factor $a$, $\delta = \delta\rho(a,\,k)/\rho(a)$ is the density contrast, $V=i\,k_jT^j_0/\rho(a)$ is the scalar velocity perturbation. The quantities $\psi$ and $\phi$ are the gravitational potentials in the Newtonian gauge. These potentials follow 
\bea
&&k^2\phi = -4\pi G a^2\sum_j \rho_j\left(\delta_j+\frac{3aH}{k^2}V_j\right)\,,\label{eqs:potential1}\\
&&k^2\left(\phi-\psi\right)= 12\pi G a^2\sum_j\left(\rho_j+p_j\right)\sigma_j\,,
\label{eqs:potential2}
\eea
where the sum runs over all the species in the Universe. We will then have sets of equations of the form of Eqs.~\eqref{eqs:perts-fluid-1} and \eqref{eqs:perts-fluid-2} depending on the number of species present in the Universe. For non-relativistic particles, i.e. cold dark matter and baryons, we just need to set $w=\delta p = \sigma=0$. However, in this paper we consider general dark energy models as well. There is no unique way to parametrize these quantities as they depend directly on the specific model considered. 

For simplicity, in this work, we consider only two components, and they are pressureless dark matter and a dark energy fluid, because we are more interested on how the different criteria reacts to a particular model. 
In the next section, we describe the different dark energy models.

Since we want to test our models with observations, we need to obtain a measurable quantity; the real observable is $f\sigma_8(a)$, defined as the product of the growth rate of matter perturbations $f(a) = \dd \ln \delta_m(a)/\dd \ln a$ and the root mean square (RMS) of matter density perturbations measured in a sphere of $8\,h^{-1}$Mpc, defined as $\sigma_8(a) = \sigma_{8,0}\delta_m (a)/\delta_m(a=1)$. We then have:
\bea
f\sigma_8(a) = \sigma_{8,0}\,a\frac{\delta'_m (a)}{\delta(a=1)}\;,
\label{definition_growth_rate_s8}
\eea
where $\sigma_{8,0}$ is the RMS measured today.  This quantity is more reliable than $f(a)$ alone due its independancy of the bias $b$, which is the ratio of baryon perturbations to total matter perturbations, i.e. $\delta_b=b\,\delta_m$.

\section{Models}\label{models}

Here we list the models considered in the analysis. Throughout this paper, we assume that all the dark models have zero anisotropic stress, $\sigma= 0$. Consequently, the two gravitational potentials are equal $\phi=\psi$. 

\subsubsection{$\Lambda$CDM}

This corresponds to the simplest and most accepted cosmological model. It assumes a constant EoS parameter exactly equal to $-1$. We consider two different cases in the $\Lambda$CDM scenario.

\paragraph*{\bf $\Lambda$CDM:} this model refers to flat $\Lambda$CDM (without spatial curvature) where we set the curvature parameter $\Omega_{k_0}=0$, hence the Hubble parameter in Eq.~\eqref{Hubble} reads
\be
H^2=H_0^2 \qty[\Omega_{m_0}a^{-3}+(1-\Omega_{m_0})]\nn\,.
\ee
Furthermore, the cosmological constant $\Lambda$ has zero perturbations, hence the system of equations simplifies and the gravitational potentials only depend on pressureless matter. 
For small scales, Eqs.~\eqref{eqs:perts-fluid-1} - \eqref{eqs:potential2} reduce to a single second-order differential equation for matter density contrast to which an analytical solution\footnote{We denote analytically-solved models by using the label `a'.} can be found, see Appendix~\ref{Append} for more details.

Consequently, we will have one model with two variants: $\Lambda$CDM and $\Lambda$CDM-a, using the numerical and analytic solution, respectively. 
However, for consistency reasons, we decided to use the full set of differential equations Eqs.~\eqref{eqs:perts-fluid-1} - \eqref{eqs:potential2}, leaving to the appendix the results obtained by using the analytical solutions as a further test. 

Finally, the parameters of both models are:
\begin{equation}
    \bm\theta_{\Lambda\text{CDM}}=(\Omega_{m_0},\,H_0,\,\sigma_{8,0})\,.
\end{equation}

\paragraph*{\bf $\Lambda$CDM-nf:} this model corresponds to a non-flat (to which we use the label `-nf') $\Lambda$CDM where we allow for the curvature parameter to vary. Then, the Hubble parameter takes the form: 
\be
H^2=H_0^2 \qty[\Omega_{m_0}a^{-3}+(1-\Omega_{m_0}-\Omega_{de_0})a^{-2}+\Omega_{de_0}]\nn\,.
\ee
The cosmological constant still has zero perturbations, however the differential equation for matter perturbations does not have an analytical solution, hence we solve numerically Eqs.~\eqref{eqs:perts-fluid-1} - \eqref{eqs:potential2}.

The parameters of the model are:
\begin{equation}
    \bm\theta_{\Lambda\text{CDM-nf}}=(\Omega_{m_0},\,\Omega_{de_0},\,H_0,\,\sigma_{8,0})\;.
\end{equation}

\subsubsection{$w$CDM}
This model is an extension of the $\Lambda$CDM model in which a constant EoS $w$ is set as a free parameter. If the EoS parameter of dark energy is no longer constant and equal to $-1$, then dark energy may have perturbations and its growth will be fully characterized by the values of $w$ and $c_s$\footnote{We remind the reader that we assume the anisotropic stress of any dark energy model to be zero.}. 
Clearly, if dark energy has perturbations, these will affect the growth of matter perturbations through the gravitational potential Eqs.~\eqref{eqs:potential1} - \eqref{eqs:potential2}. 
We identify four different cases.

\paragraph*{\bf  $w$CDM:} this model corresponds to flat  $w$CDM where perturbations in the dark energy sector have been switched off; the Hubble parameter reads:
\begin{equation}
    H^2=H_0^2 \qty[\Omega_{m_0}a^{-3}+(1-\Omega_{m_0})a^{-3(1+w)}]\,.
    \label{hubble_parameter_wcdm}
\end{equation}
If we decide to ignore \emph{a priori} the dark energy perturbations, then the growth of matter density is still governed by a second order differential equation, and it is still possible to find an analytical solution to the matter density contrast, see Appendix~\ref{Append} for more details. As a consequence we have the $w$CDM and $w$CDM-a solutions to this model.
%\cite{Belloso:2011ms} 
%\bea
%\delta(a)=a_2F_1\bigg(&&\frac{w-1}{2w},\,-\frac{1}{3w},\,1-\frac{5}{6w},\nn\\
%&&1-\Omega_m^{-1}(a)\bigg)\;.
%\eea
As for the $\Lambda$CDM case, we also consider the full numerical solutions from Eqs.~\eqref{eqs:perts-fluid-1} - \eqref{eqs:potential2} and leaving the results from the analytical solution to the appendix. 

Finally, the parameters of both models are
\begin{equation}
    \bm\theta_{\text{wCDM}}=(\Omega_{m_0},\,w,\,H_0,\,\sigma_{8,0})\;.
\end{equation}

\paragraph*{\bf $w$CDM-nf:} this model corresponds to a non-flat  $w$CDM; the Hubble parameter reads
\bea
    H^2=H_0^2 \Big[&&\Omega_{m_0}a^{-3}+(1-\Omega_{m_0}-\Omega_{de_0})a^{-2}+ \nn \\
      &&\Omega_{de_0}a^{-3(1+w)}\Big]\,.
    \label{hubble_parameter_wcdm_nonflat}
\eea 
Here we set dark energy perturbations to zero. However, due to the complexity of the Hubble parameter, analytical solutions for the matter density contrast do not exist and we solve numerically the system of Eqs.~\eqref{eqs:perts-fluid-1} - \eqref{eqs:potential2}.

We have the following free parameters for the model:
\begin{equation}
    \bm\theta_{\text{wCDM-nf}}=(\Omega_{m_0},\, \Omega_{de_0},\,w, \,H_0,\,\sigma_{8,0})\;.
\end{equation}
 
\paragraph*{\bf $w$CDM-p:} this model is a flat $w$CDM for which we allow perturbations (this addition is symbolized by `-p') in the dark energy sector. The Hubble parameter is given by Eq.~\eqref{hubble_parameter_wcdm}. However, we now have two sets of equations \eqref{eqs:perts-fluid-1} - \eqref{eqs:perts-fluid-2}, for pressureless matter and for the dark energy fluid. Analytical solutions can also be found in some special limits, see Appendix \ref{Append}. 
% in which case the density contrast is given by
%\bea
%\delta(a)=a_2F_1\bigg(\frac{1}{4}&-&\frac{5}{12w}+B,\frac{1}{4}-\frac{5}{12w}-B,\nn\\
%1&-&\frac{5}{6w},1-\Omega_m^{-1}(a)\bigg)
%\eea
%where $B$ is used as $B_\text{joint}$ in \cite{Nesseris:2015fqa}, which corresponds to:
%\bea
%B=\frac{1}{12w}\sqrt{(1-3w)^2+24\frac{1+w}{1-3w+\frac{2}{3}\frac{k^2}{H_0^2\Omega_{m_0}}c_s^2}}\;.
%\eea
However, as for the other cases, we also use the full numerical solutions from the equation of perturbations. 

As mentioned earlier, the growth of the perturbations of one species depends on the characteristics of the fluid, which are given by $w$, $\delta p$ and $\sigma$. For pure pressureless matter, $w=\delta p = \sigma=0$. For a dark energy fluid, we assume zero anisotropic stress $\sigma=0$, and the pressure perturbation to be given by \cite{Kunz:2006wc}:
\begin{equation}
\delta p = c_s^2\rho\delta+\frac{3aH(c_s^2-c_a^2)}{k^2}\rho V\;,
\label{eq:dp-darkenergy}
\end{equation}
where $c_a^2\equiv\dot{p}/\dot{\rho}$ is the adiabatic sound speed of the fluid that can be expressed as 
\begin{equation}
c_a^2=w-\frac{\dot{w}}{3H(1+w)}=w-\frac{w'}{3(1+w)}\,,
\end{equation}
and for a constant EoS, the adiabatic sound speed becomes $c_a^2=w$.

The free parameters of the models ($w$CDM-p and $w$CDM-p-a) are
\begin{equation}
    \bm\theta_{\text{wCDM-p}}=(\Omega_{m_0},\,w,\,c_s^2,\,H_0,\,\sigma_{8,0})\;.
\end{equation}

\paragraph*{\bf $w$CDM-nf-p:} this model corresponds to a non-flat $w$CDM for which we allow perturbations in the dark energy sector; the Hubble parameter takes the form in Eq.~\eqref{hubble_parameter_wcdm_nonflat} and the perturbations will be solved numerically for both matter and dark energy. 
Thus, the parameter set of the model is
\begin{equation}
    \bm\theta_{\text{wCDM-nf-p}}=(\Omega_{m_0},\,\Omega_{de_0},\,w,\,c_s^2,\,H_0,\,\sigma_{8,0})\;.
\end{equation}

\subsubsection{Chevallier-Polarski-Linder (CPL)}
This class of models \cite{Chevallier:2000qy, Linder:2002et} can be considered an extension to $w$CDM models in which the equation of state depends on the scale factor. The simplest extension is a Taylor expansion around the present time $a=1$, giving
\begin{equation}
    w(a)=w_0 + w_a (1-a)\,.
    \label{wCPL}
\end{equation}
Hence, giving two extra parameters: $w_0$, which is the present time EoS parameter and $w_a$ which represents the variation over time of $w(a)$. We identify four different models using this parametrization.

\paragraph*{\bf CPL:} this corresponds to the simplest scenario where the Hubble parameter does not depend on curvature and we set dark energy perturbations to zero. Then, the Hubble parameter reads
\begin{equation}
    H^2=H_0^2 \qty[\Omega_{m_0}a^{-3}+(1-\Omega_{m_0})a^{-3(1+\hat{w}(a))}]\,. 
    \label{hubble_parameter_cpl}
\end{equation}
There is no exact analytic expression for the matter density contrast when the EoS parameter takes the form of Eq.~\eqref{wCPL} but only approximated analytical solutions, \cite{Belloso:2011ms}. 
Here we solve numerically Eqs.~\eqref{eqs:perts-fluid-1} - \eqref{eqs:potential2}. 
This way, the parameters of are
\begin{equation}
    \bm\theta_{\text{CPL}}=(\Omega_{m_0},\,w_0,\,w_a, \,H_0,\,\sigma_{8,0})\;.
\end{equation}
\paragraph*{\bf CPL-nf:} in this model we allow the curvature parameter to vary. Then, Hubble parameter becomes
\bea
 H^2=H_0^2 && \left[\Omega_{m_0}a^{-3}+(1-\Omega_{m_0}-\Omega_{de_0})a^{-2}+\right. \nn \\
     && \left. \Omega_{de_0}a^{-3(1+\hat{w}(a))}\right]\,.
    \label{hubble_parameter_cpl_nonflat}
\eea
We set dark energy perturbations to zero and solve numerically the Eqs.~\eqref{eqs:perts-fluid-1} - \eqref{eqs:potential2} for pure pressureless matter only. 
The free parameters of the model are:
\begin{equation}
    \bm\theta_{\text{CPL-nf}}=(\Omega_{m_0},\,\Omega_{de_0},\,w_0,\,w_a, \,H_0,\,\sigma_{8,0})\;.
\end{equation}

\paragraph*{\bf CPL-p:} the Hubble parameter is given by Eq.~\eqref{hubble_parameter_cpl}, the equation of perturbations will be solved numerically by using Eqs.~\eqref{eqs:perts-fluid-1} - \eqref{eqs:potential2}. 
The characteristics of the dark energy fluid are given by Eq.~\eqref{eq:dp-darkenergy}, with the further assumption that the adiabatic sound speed $c_a^2=w$; the former is somehow required in order to stabilize the growth of dark energy perturbations when it crosses the phantom regime \cite{Kunz:2006wc}.
Thus, the parameters of the model are
\begin{equation}
    \bm\theta_{\text{CPL-p}}=(\Omega_{m_0},\,w_0,\,w_a,\,c_s^2,\,H_0,\,\sigma_{8,0})\;.
\end{equation}

\paragraph*{\bf CPL-nf-p:} the Hubble parameter takes the form in Eq.~\eqref{hubble_parameter_cpl_nonflat}. We solve numerically Eqs.~\eqref{eqs:perts-fluid-1} - \eqref{eqs:potential2} for pressureless matter and dark energy. 
The characteristic of the dark energy fluid are given by Eq.~\eqref{eq:dp-darkenergy}, with the further assumption of $c_a^2=w$ and $w'=0$ at crossing.
\\
The parameter set for the model is
\begin{equation}
    \bm\theta_{\text{CPL-nf-p}}=(\Omega_{m_0},\,\Omega_{de_0},\,w_0,\,w_a,\,c_s^2, \,H_0,\,\sigma_{8,0})\;.
\end{equation}

\section{Data}\label{data}

The Hubble parameter data for the analysis are the \emph{cosmic chronometers} compilation used in~\cite{Marra:2017pst}, which consists in 31 independent measurements of $H(z)$, obtained from evolving galaxies at different redshifts~\cite{Moresco:2016mzx}.

\begin{table}[H]%[htp]
\begin{center}
\begin{tabular}{ccccccccc}
\hline
\hline
$z$ & $H(z)$ & $\sigma_{H}(z)$& Ref. & \phantom{jhmes}&$z$ & $H(z)$ & $\sigma_{H}(z)$ & Ref.\\
\hline
 0.07 & 69.0 & 19.6 & \cite{Zhang:2012mp} & &0.4783 & 80.9 & 9.0  & \cite{Moresco:2016mzx} \\
 0.09 & 69.0 & 12.0 & \cite{Simon:2004tf} & & 0.48 & 97.0 & 62.0 & \cite{Stern:2009ep}  \\
 0.12 & 68.6 & 26.2 & \cite{Zhang:2012mp}  & & 0.593 & 104.0 & 13.0 & \cite{Moresco:2012jh} \\
 0.17 & 83.0 & 8.0 & \cite{Simon:2004tf}  & & 0.68 & 92.0 & 8.0 & \cite{Moresco:2012jh} \\
 0.179 & 75.0 & 4.0 & \cite{Moresco:2012jh}  & & 0.781 & 105.0 & 12.0  & \cite{Moresco:2012jh} \\
 0.199 & 75.0 & 5.0 & \cite{Moresco:2012jh}  & & 0.875 & 125.0 & 17.0 & \cite{Moresco:2012jh} \\
 0.2 & 72.9 & 29.6 & \cite{Zhang:2012mp}  & & 0.88 & 90.0 & 40.0 & \cite{Stern:2009ep} \\
 0.27 & 77.0 & 14.0 & \cite{Simon:2004tf}  & & 0.9 & 117.0 & 23.0 & \cite{Simon:2004tf} \\
 0.28 & 88.8 & 36.6 & \cite{Zhang:2012mp}  & & 1.037 & 154.0 & 20.0 & \cite{Moresco:2012jh} \\
 0.352 & 83.0 & 14.0 & \cite{Moresco:2012jh}  & & 1.3 & 168.0 & 17.0 & \cite{Simon:2004tf} \\
 0.3802 & 83.0 & 13.5  & \cite{Moresco:2016mzx} & & 1.363 & 160.0 & 33.6  & \cite{Moresco:2015cya} \\
 0.4 & 95.0 & 17.0 & \cite{Simon:2004tf}  & & 1.43 & 177.0 & 18.0 & \cite{Simon:2004tf} \\
 0.4004 & 77.0 & 10.2 & \cite{Moresco:2016mzx}  & & 1.53 & 140.0 & 14.0 & \cite{Simon:2004tf} \\
 0.4247 & 87.1 & 11.2  & \cite{Moresco:2016mzx} & & 1.75 & 202.0 & 40.0 & \cite{Simon:2004tf} \\
 0.4497 & 92.8 & 12.9  & \cite{Moresco:2016mzx} & & 1.965 & 186.5 & 50.4  & \cite{Moresco:2015cya} \\
 0.47 & 89.0 & 49.6  & \cite{Ratsimbazafy:2017vga} & & & & & \\
\hline
\hline
\end{tabular}
\end{center}
\caption{The 31 cosmic chronometer data points used in this analysis along with their related references. The $H(z)$ and $\sigma_{H}(z)$ data are in units of km\,s$^{-1}$\,Mpc$^{-1}$.}
\label{tab:tableH}
\end{table}

The growth rate dataset is based on the compilation used in~\cite{Sagredo:2018ahx}, which is an updated version of the `Gold-2017' dataset from~\cite{Nesseris:2017vor}. The dataset consists of 22 independent measurements of $f\sigma_8(z)$, obtained through baryon acoustic oscillations and weak lensing surveys. Among these surveys, it is important to note that the three WiggleZ~\cite{Blake:2012pj} and the four SDSS-IV~\cite{Zhao:2018jxv} measurements are correlated, and their covariance matrices are
\begin{equation}\label{WiggleZCov}
    \vb C_{\text{WiggleZ}}= 10^{-3}
    \mqty(6.400 & 2.570 & 0.000 \\
    2.570 & 3.969 & 2.540 \\
    0.000 & 2.540 & 5.184)\;,
\end{equation}
\begin{equation}\label{SDSS4Cov}
    \vb C_{\text{SDSS-IV}}= 10^{-2}
    \mqty(3.098 & 0.892 &  0.329 & -0.021\\
       0.892 & 0.980 & 0.436 & 0.076\\
       0.329 & 0.436 &  0.490   & 0.350 \\
       -0.021 & 0.076 & 0.350 & 1.124) \;.
\end{equation}

\section{Methodology}\label{methodology}

To perform the analysis, both datasets are assumed to have Gaussian likelihood distributions, this is the probability of the data given a set of parameters. The datasets are assumed to be independent, thus their conjoined likelihood is the product of each dataset's likelihood. In terms of the traditional chi-squared, defined by $\chi^2\equiv -2\log{L}$, where $L$ is the likelihood of the current model, it is simply given by the sum of each dataset's chi-squared, or
\begin{equation}
\chi^2  =\chi^2_{H}+ \chi^2_{f\sigma_8}
\end{equation}
Where the subscripts `$f\sigma_8$' and `$H$' indicate growth and expansion contributions, respectively.

\begin{table}[H]%[htp]
    \begin{center}
        \begin{tabular}{ccccccccc}
            \hline
            \hline
            $z$     & $f\sigma_8(z)$ & $\sigma_{f\sigma_8}(z)$  & $\Omega_{m_0}^\text{ref}$ & Ref. \\ \hline
            0.02    & 0.428 & 0.0465  & 0.3 & \cite{Huterer:2016uyq}   \\
            0.02    & 0.398 & 0.065   & 0.3 & \cite{Turnbull:2011ty},\cite{Hudson:2012gt} \\
            0.02    & 0.314 & 0.048   & 0.266 & \cite{Davis:2010sw},\cite{Hudson:2012gt}  \\
            0.10    & 0.370 & 0.130   & 0.3 & \cite{Feix:2015dla}  \\   
            0.15    & 0.490 & 0.145   & 0.31 & \cite{Howlett:2014opa}  \\   
            0.17    & 0.510 & 0.060   & 0.3 & \cite{Song:2008qt}  \\  
            0.18    & 0.360 & 0.090   & 0.27 & \cite{Blake:2013nif} \\  
            0.38    & 0.440 & 0.060   & 0.27 & \cite{Blake:2013nif} \\ 
            0.25    & 0.3512 & 0.0583 & 0.25 & \cite{Samushia:2011cs} \\ 
            0.37    & 0.4602 & 0.0378 & 0.25 & \cite{Samushia:2011cs} \\ 
            0.32    & 0.384 & 0.095  & 0.274 & \cite{Sanchez:2013tga}   \\
            0.59    & 0.488  & 0.060 & 0.307115 & \cite{Chuang:2013wga} \\
            0.44    & 0.413  & 0.080 & 0.27 & \cite{Blake:2012pj} \\
            0.60    & 0.390  & 0.063 & 0.27 & \cite{Blake:2012pj} \\
            0.73    & 0.437  & 0.072 & 0.27 & \cite{Blake:2012pj} \\
            0.60    & 0.550  & 0.120 & 0.3 & \cite{Pezzotta:2016gbo} \\
            0.86    & 0.400  & 0.110 & 0.3 & \cite{Pezzotta:2016gbo} \\
            1.40    & 0.482  & 0.116 & 0.27 & \cite{Okumura:2015lvp} \\
            0.978   & 0.379  & 0.176 & 0.31 & \cite{Zhao:2018jxv} \\
            1.23    & 0.385  & 0.099 & 0.31 & \cite{Zhao:2018jxv} \\
            1.526   & 0.342  & 0.070 & 0.31 & \cite{Zhao:2018jxv} \\
            1.944   & 0.364  & 0.106 & 0.31 & \cite{Zhao:2018jxv} \\
            \hline
            \hline
        \end{tabular}
    \end{center}
    \caption{Compilation of the cosmic growth $f\sigma_8(z)$ measurements used in this analysis along with the reference matter density parameter $\Omega_{m_0}$ (needed for the redshift correction) and associated references.}
    \label{tab:tablefs8}
\end{table}

Let us suppose that there are $n$ measurements of $H$ or $f\sigma_8$, so we represent the observed data in different redshifts as $\vb m=(m(z_1),\dots, m(z_n))$ and its theoretical prediction as $\bm \mu(\bm \theta)=(\mu(z_1),\dots, \mu(z_n))$, which depend on the cosmological model and parameters. We define the data vector as
\begin{equation}
    \vb x_s = \vb m_s - \bm\mu_s\;,
\end{equation}
with the subscript `$s$' denoting the data source: $H$ or $f\sigma_8$. However, in the case of growth measurements, we need to take into account a redshift correction, which is featured in Ref.~\cite{Nesseris:2017vor}. This correction consists in the following factor
\begin{equation}
\text{fac}(z_i)= \frac{H(z_i)d_A(z_i)}{H^{\text{ref},i}(z_i) d_A^{\text{ref},i}(z_i)}\,
\end{equation}
where the superscript `$\text{ref},i$' indicates that the reference cosmology is taken on the corresponding data point at redshift $z_i$. With this procedure, we arrive at the corrected growth theoretical prediction:
\begin{equation}
\mu_{c}^i=\frac{\mu^i_{f\sigma_8}}{\text{fac}(z_i)}\,.
\end{equation}
For all the datapoints, the reference model used is $\Lambda$CDM, and one can note that the product $H(z)d_A(z)$ is independant of $H_0$ and $\sigma_{8,0}$ for all models considered. We list the reference values for $\Omega_{m_0}$ of each datapoint in Table~\ref{tab:tablefs8}. Using the corrected prediction, the data vector for $f\sigma_8$ is
\begin{equation}
     \vb x_{f\sigma_8} = \vb m_{f\sigma_8} - \bm\mu_{c}\;.
\end{equation}
Therefore, the chi-squared are constructed through
\begin{equation}
    \chi^2_{s}= \vb x_{s}^T \vb C_{s}^{-1} \vb x_{s}\;,
\end{equation}
where $\vb C_s^{-1}$ the inverse of the covariance matrix of the dataset. In the case of cosmic expansion, the covariance matrix is diagonal and equal to each datapoint's variance. The total cosmic growth covariance matrix is given by a diagonal matrix with the measurements' variance, with the insertion of the WiggleZ matrix and SDSS-IV matrices, given by Eqs.~\eqref{WiggleZCov} and \eqref{SDSS4Cov}.

\begin{table}[htp]
\begin{center}
\begin{tabular}{cc}
\hline
\hline
Parameter &  Flat prior limits \\ 
\hline
$\Omega_{m_0}$ & $[0, 1]$ \\
$\Omega_{de_0}$ & $[0, 1.7]$ \\
$w_0$ & $[-3.5, -1/3]$ \\
$w_a$ & $[-2.5, -1/3-w_0]$ \\
$c_s^2$ & $[0, 1]$ \\
$H_0$ [Mpc/km/s] & $[35, 110]$ \\
$\sigma_{8,0}$ & $[0.3, 1.5]$ \\
\hline
\hline
\end{tabular}
\end{center}
\caption{Ranges of the flat priors used for each parameter. Note that $w_a$ depends on the value of $w_0$ to define its upper bound. This is to ensure that $w(a)<-1/3$ in order to have acceleration on the expansion of the Universe.}
\label{tab:priors}
\end{table}

We now proceed to present the methods used to compare different dark energy models. We use five methods in total.

\paragraph*{\bf Evidence.} The first method is the standard Bayesian model comparison via evidence computation $\log(E)$ \cite{Trotta:2005ar}, where the evidence is defined via
\begin{equation}
E(\vb m|\,M) =\int L(\vb m|\,\bm \theta_{M},M)\,\pi(\bm \theta_M|\,M)\,\d \bm\theta_M\;.
\end{equation}
The former quantity determines the probability of a given model $M$ to be true, given the data $\vb m$. As already mentioned, the likelihood function $L(\vb m |\,\bm \theta_{M},M)$ is Gaussian on the data $\vb m$, and the prior probability for the parameters, $\pi(\bm \theta_M|\,M)$. If we assume the prior probabilities $\pi(M)$ to be the same for each model, then the Evidence completely defines the ranking of the cosmological models. 

All throughout the analysis, and specifically for the evidence computation, we adopted standard flat priors for all the parameters, with boundaries reported in Table~\ref{tab:priors}. Despite that, the only special treatment was made on $w_a$, for which we used an upper bound that depends on the value of $w_0$ in order to guarantee a phase of accelerated expansion \cite{Chevallier:2000qy}. Furthermore, we use the same priors for all models that have the free parameter, as we are more concerned about the statistical methods used.

The computation is performed using the package Nestle \cite{Nestle}, a Python implementation of the MultiNest algorithm \cite{Shaw:2007jj,Feroz:2008xx}. This algorithm is an efficient and robust way of computing the evidence integral, a numeric task that becomes too large to be grid-integrated. MultiNest also produces a Markov chain that can be reused as the MCMC-sample for the next method below.

\paragraph*{\bf Figure of Merit.} With this method, the models are ranked by their FoM's defined in \cite{Basilakos:2017rgc}, which corresponds to the inverse of the $1\sigma$ confidence region area in the conjoined $f\sigma_8-H$ plot given a redshift range. The likelihood is used to MCMC-sample in the parameter space of each model, and this parameter chain is used to get the $1\sigma$ range of $f\sigma_8(z_i)$ for $i \in {1,...,n}$. If there are sufficient $z_i$ points, a spline can be constructed to connect the points in the $f\sigma_8-H$ plane, keeping $H(z_i)$ fixed to its mean value. This method is viable because $f\sigma_8$ is much less constrained than $H$ in all the models tested, and $H(z)$ increases monotonically with $z$ for each model. The redshift range is, in principle, defined between $z=0$ to $z_{max}=2$, to include the whole redshift data range. We will also show how the FoM varies when $z_{max}$ changes.

\begin{figure}[H]
 %   \centering
    \includegraphics[width=0.47\textwidth]{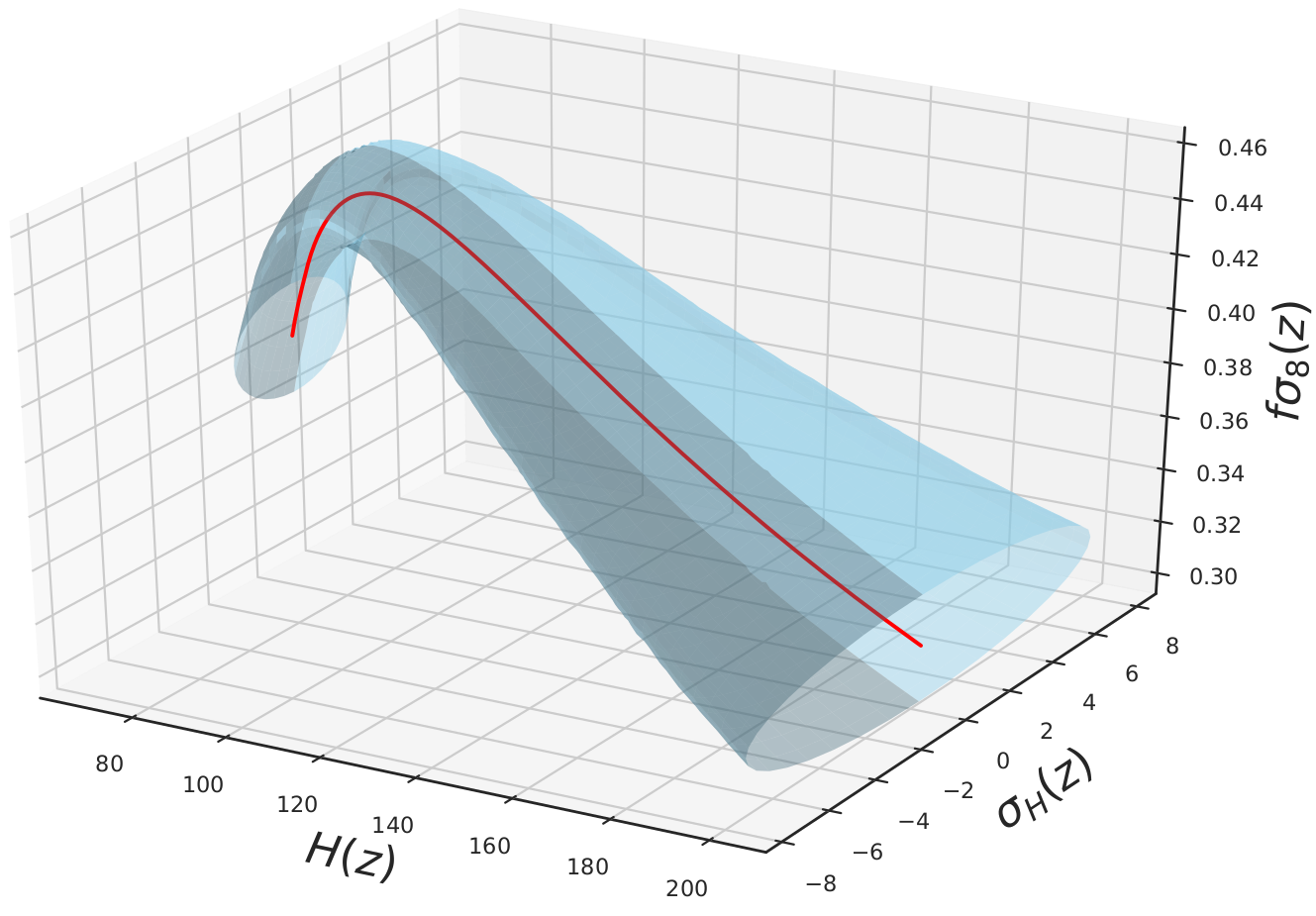}
    \caption{3-FoM plot.}
    \label{fig:3-FoM-plot}
\end{figure}

\paragraph*{\bf 3-FoM.} Here, we propose an extension of the previous method, in which we now consider the $1\sigma$ range of $H(z_i)$ (as opposed as in the last method where it was omitted). For a $z_i$ point we obtain the values plus the associated confidence levels of the Hubble parameter and the growth rate, i.e.
\be 
H(z_i)^{+\sigma_{H(z_i)_{+}}}_{-\sigma_{H(z_i)_{-}}} \quad \text{and} \quad f\sigma_8(z_i)^{+\sigma_{f\sigma_8(z_i)_{+}}}_{-\sigma_{f\sigma_8(z_i)_{-}}}\,.\nn 
\ee
With these values we compute the ellipsoidal area on each redshift point $z_i$, as an approximation for the 2-dimensional confidence region in the $f\sigma_8(z_i), H(z_i)$ space.
\be
A_e(z_i)=\frac{\pi}{4}\qty(\sigma_{H(z_i)_{+}}+\sigma_{H(z_i)_{-}})
\qty(\sigma_{f\sigma_8(z_i)_{+}}+\sigma_{f\sigma_8(z_i)_{-}})\,.\nn
\ee
The 3-FoM is defined as the inverse of the ellipsoidal volume quantity in the $f\sigma_8, H, \sigma_H$ space, being 
\begin{align}
V_e&=\int A_e(z)d H(z)\nn \\
&=-\int_{z=0}^{z=2}A_e(z)\frac{H'(z)}{(1+z)^2}dz\nn \\
&\simeq -\sum_i A_e(z_i)\frac{H'(z_i)}{(1+z_i)^2}\Delta z\,.
\end{align}
If there are many equispaced $z_i$ points, the previous quantity corresponds to the volume enclosed in Fig.~\ref{fig:3-FoM-plot}.

\paragraph*{\bf BIC.} The fourth method is the Bayesian Information Criterion \cite{Schwarz:1978tpv,Liddle:2004nh} which is given by:
\begin{equation}
\text{BIC}= 2 \ln(N_\text{data})n_\text{pars} - 2 \ln{L_\text{max}}\,.
\end{equation}
This method still considers the maximum likelihood $L_\text{max}$, however it tends to penalize models with several parameters through the direct dependence of $n_\text{pars}$ . 
Its formulation aims at approximating the evidence (specifically, $-2\log(E)$) of the model to be tested, hence the favored model is the one with the lowest BIC value.

\paragraph*{\bf $\text{AIC}_c$.} The last statistical method is the corrected Akaike Information Criterion ($\text{AIC}_c$)  \cite{AIC2}. This method is similar to the BIC method because it still penalizes models with several parameters, however the penalisation is weighted with the number of data. Contrary to the BIC test, the $\text{AIC}_c$ tends to favor one model if the data set is large enough. The criterion is given by:
\begin{equation}
\text{AIC}_c= 2 n_\text{pars} - 2 \ln{L_\text{max}} + \frac{n_\text{pars}(n_\text{pars}+1)}{N_\text{data}-n_\text{pars}-1}\,.
\end{equation}
This equation, derived in \cite{AIC2}, accounts for a correction term when the number of data is small, unlike the original Akaike Information Criterion \cite{AIC}. As before, the test should also be similar to the value of $-2\log(E)$, which means that the lower $\text{AIC}_c$ is, the more favored is the model. 

\begin{figure}%[H]
 %   \centering
    \includegraphics[width=0.48\textwidth]{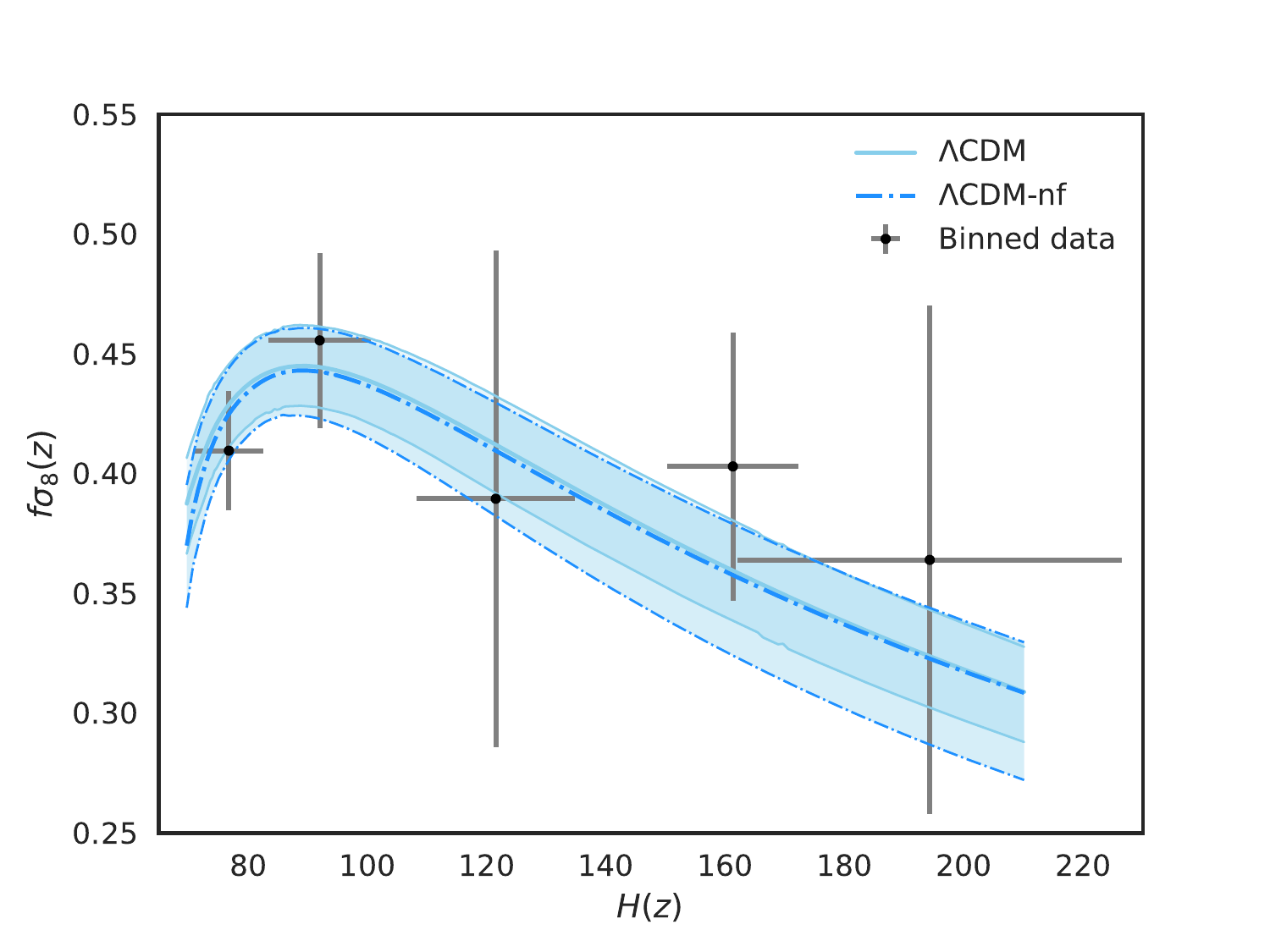}\\
     \includegraphics[width=0.48\textwidth]{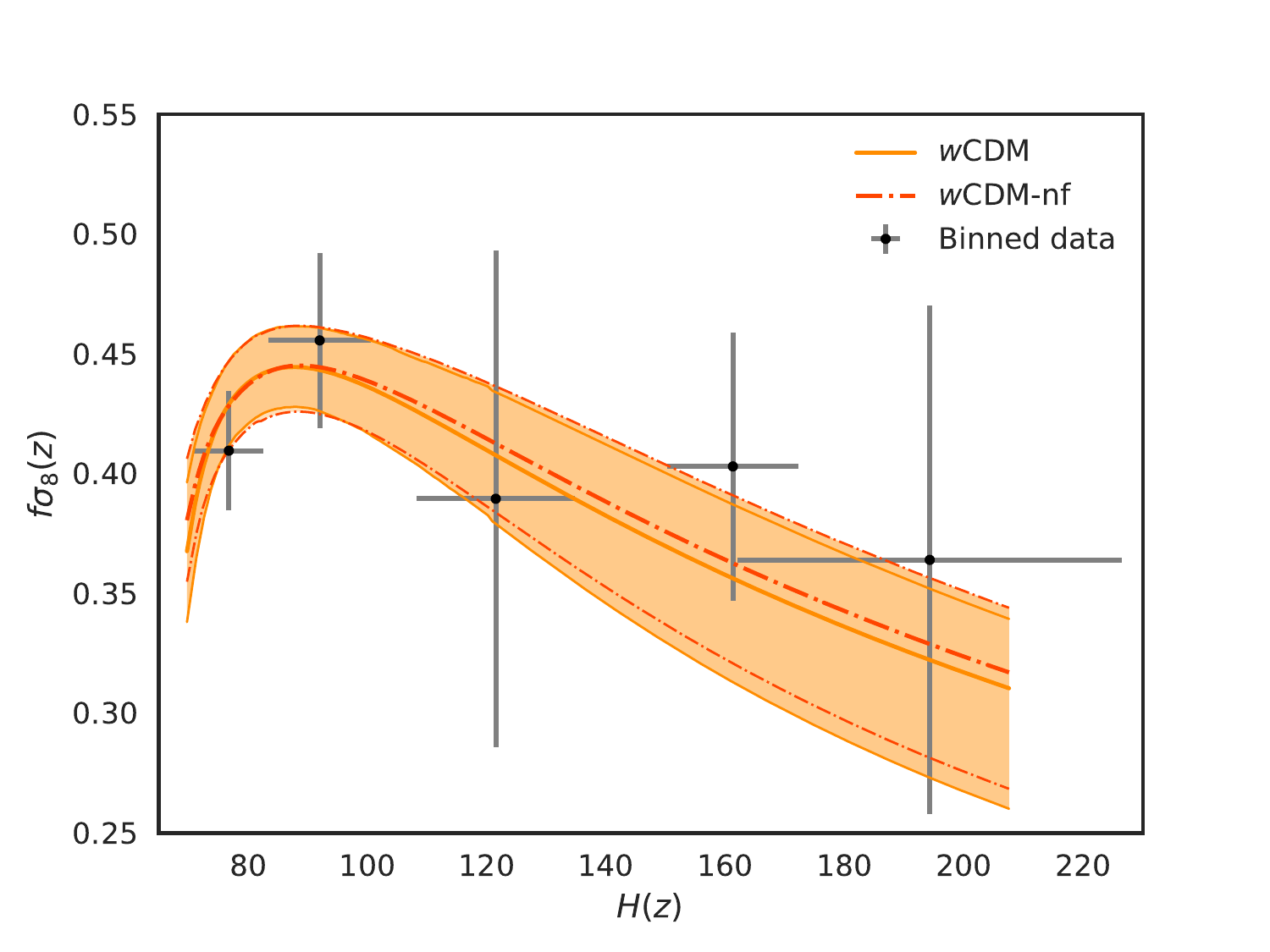}\\
    \includegraphics[width=0.48\textwidth]{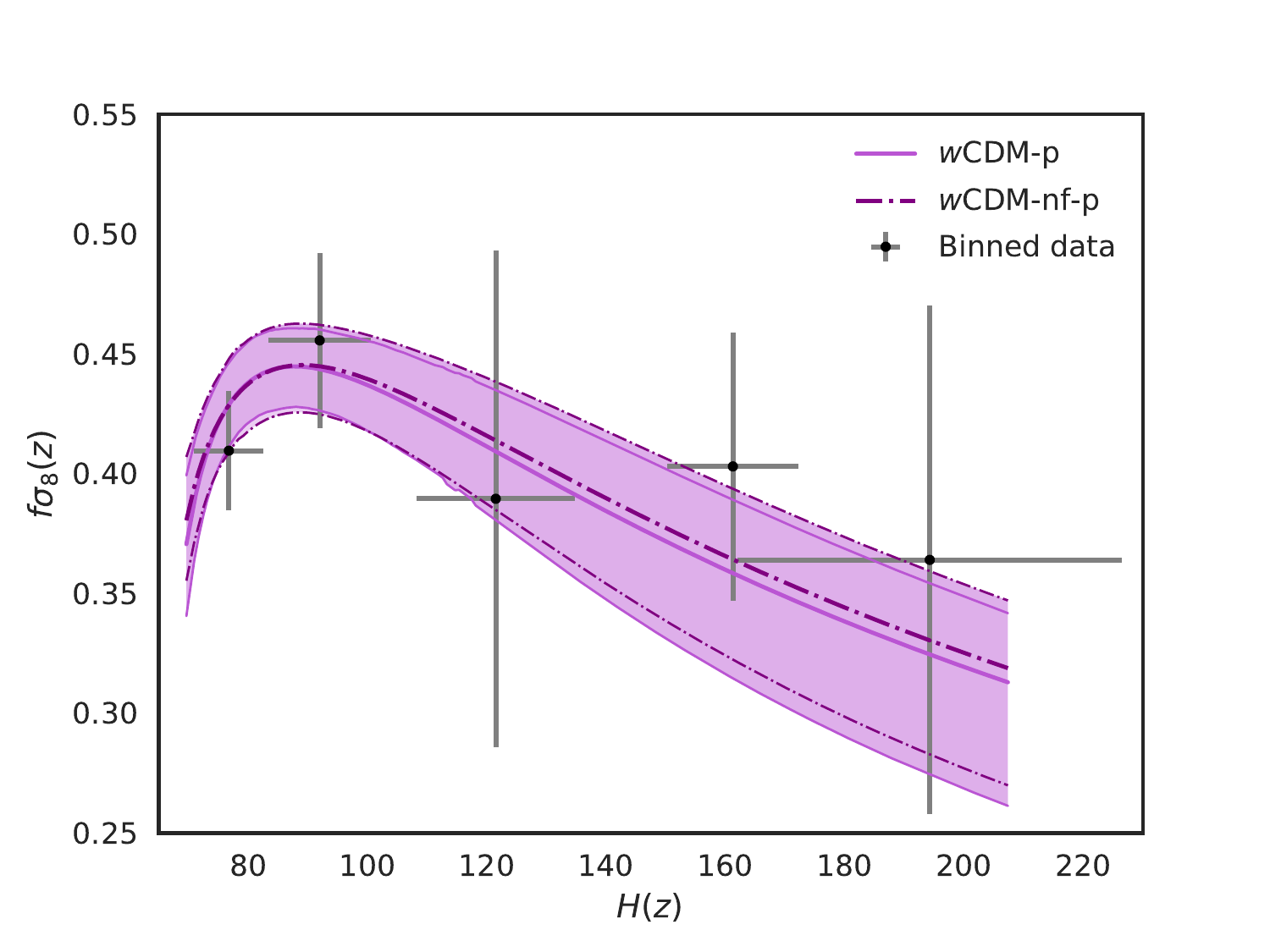}\\
    \caption{The conjoined plots of the cosmic growth $f\sigma_8(z)$ versus the cosmic expansion $H(z)$ for differents models described in the text: (upper panel) $\Lambda$CDM with $\Lambda$CDM-nf, (middle panel) $w$CDM with $w$CDM-nf and (lower panel) $w$CDM-p with $w$CDM-nf-p. Also the 1$\sigma$ error regions (shaded areas) and the real binned data (gray points) are shown.}
    \label{figs:L-w-joint}
\end{figure}

\section{Results and Discussion}\label{results}

In this section, we discuss the results found for each model and we compare the values of the criteria used. As mentioned previously the goal of the paper is to accurately test the common criteria found in literature and to highlight their differences. 

In general, we are not interested in the specific value of the criterion found for a particular model but rather their difference between two models. 
This difference will tell us which is the model that is able to better reproduce the data. 

For the first criterion, i.e. the evidence $E$, we use Jeffrey's scale which is defined as the difference of the logarithmic evidences for two particular models, we report it in Tab.~\ref{Jeffreys} for completeness. 

The other two criteria, i.e. BIC and AIC$_c$, are directly connected to the likelihood of the models and hence they can be used as model selection tests. Since they come from a Taylor expansion around the maximum likelihood estimator of the likelihood function, they can be connected to Jeffrey's scale, however, this interpretation must be taken with care, see \cite{Nesseris:2012cq} for a detailed discussion. 
Generally, we can still consider the difference $\left|\Delta \textrm{BIC}\right| = \left|\textrm{BIC}_2-\textrm{BIC}_1\right|$, where the index $2$ refers to the model with the higher value of BIC and the index $1$ to the one with the lower, as a good model selection test. 
Specifically, if $\left|\Delta \textrm{BIC}\right|\leq 2$, then there is no evidence in support of a model, if $2<\left|\Delta \textrm{BIC}\right|\leq 6$, then there is a positive evidence in favor of the model with the smaller value, whereas if $\left|\Delta \textrm{BIC}\right| > 6$, the evidence is considered to be strong. 
The same discussion applies to the AIC$_c$ criterion, where in this case we have: if the difference is less than $2$, then both models are able to reproduce the data with the same accuracy, if $\left|\Delta \textrm{AIC}\right|$ is between $2$ and $4$, then there is a positive evidence for the model with the lower AIC$_c$, instead if $\left|\Delta \textrm{AIC}\right|>10$, then the model with the larger AIC$_c$ is strongly disfavored, see \cite{Perez-Romero:2017njc}.

The last two criteria considered in this work are the FoM, defined as the inverse of the enclosed area at $1\sigma$ level for $f\sigma_8(z)$, and the 3-FoM defined as the inverse of the enclosed volume at $1\sigma$ level in both $f\sigma_8(z)$ and $H(z)$. It is clear the FoM and its extension (3-FoM) are not criteria able to favor/disfavor a model, but rather they give an estimation on the sensitivity of the parameters according to the data used. In practice, a larger FoM and/or 3-FoM means that the model is better constrained by the data. 

Fig.~\ref{figs:L-w-joint} (top panel) shows the reconstruction of the $H(z)-f\sigma_8(z)$ assuming flat and non-flat $\Lambda$CDM as the cosmological model. The shaded areas are obtained directly from the 1$\sigma$ errors of the parameters given by the MCMC samples. 
The best fit of the parameters are reported in Tab.~\ref{tab:par-constraints}. For this particular model the addition of an extra parameter, $\Omega_{de_0}$, alters the results and the two shaded areas differ, specially at high redshift where the lower limit of the errors are larger for not flat $\Lambda$CDM: as a consequence the FoM and 3-FoM decrease of about $35\%$ and $125\%$, respectively. The BIC and AIC$_c$ criteria used in this analysis increase of about $13\%$ and $5\%$ when the curvature parameter is considered, see Tab.~\ref{tab:results}. 
 
As for the model comparison, the evidence gives inconclusive results, the AIC$_c$ criterion favors positively the flat $\Lambda$CDM over the non-flat $\Lambda$CDM model, the BIC criterion instead shows a strong evidence in favor of $\Lambda$CDM.

\begin{table}[H]
\begin{center}
\begin{tabular}{lcccccc}
\hline
\hline
Model &  $\log(E)$ &    FoM & 3-FoM &    BIC &    AIC$_c$ & $H_{max}$\\
\hline
$\Lambda$CDM      &   -21.87 &  0.192 & 0.027 &  51.35 &  34.02 &  201.71 \\
$\Lambda$CDM-nf   &   -22.09 &  0.145 & 0.012 &  58.84 &  35.91 &  210.06 \\
$w$CDM               &   -23.50 &  0.124 & 0.013 &  59.31 &  36.39 &  204.89 \\
$w$CDM-p            &   -23.01 &  0.124 & 0.014 &  67.21 &  38.78 &  204.47 \\
$w$CDM-nf           &   -23.33 &  0.125 & 0.010 &  66.57 &  38.14 &  207.54 \\
$w$CDM-nf-p          &   -23.29 &  0.122 & 0.010 &  74.80 &  40.99 &  207.36 \\
CPL                &   -24.16 &  0.129 & 0.014 &  67.18 &  38.75 &  208.97 \\
CPL-p               &   -24.14 &  0.127 & 0.014 &  74.87 &  41.05 &  208.69 \\
CPL-nf              &   -24.53 &  0.119 & 0.010 &  74.51 &  40.69 &  204.65 \\
CPL-nf-p            &   -24.53 &  0.120 & 0.010 &  82.62 &  43.52 &  204.33 \\
\hline
\hline
\end{tabular}
\end{center}
\caption{Results of the different methods for each model. We also show $H_{max}=H(z=2)$ to compare the extension of the integration in the $H$-dimension for the FoM and 3-FoM methods.} 
\label{tab:results}
\end{table}

\begin{table}[h]
    \caption{Jeffrey's Scale as in Ref.~\cite{Trotta:2008qt}, which compares the logarithmic Evidence difference between the two models. The different levels represent different degrees of belief in that one is the true theory.}
    \begin{center}
        \begin{tabular}{ccc}
            \hline
            \hline
            $ |\Delta\log(E)|$ & Probability  & Evidence\\ \hline
            $0\leq |\Delta\log(E)| <1.0$ & $0\leq P_1<0.75$ & Inconclusive\\
            $1.0\leq |\Delta\log(E)| <2.5$ & $0.75\leq P_1<0.923$ &  Weak\\
            $2.5\leq |\Delta\log(E)| <5.0$ & $0.923\leq P_1<0.993$ & Moderate\\
            $5.0\leq |\Delta\log(E)|$ & $0.993\leq P_1$ & Strong\\
            \hline
            \hline
        \end{tabular}
    \end{center}
    \label{Jeffreys}
\end{table}

In Fig.~\ref{figs:L-w-joint} (middle panel)  are shown the  reconstruction of the $H(z)-f\sigma_8(z)$ assuming flat and non-flat $w$CDM. These models have one parameter more with respect to the corresponding $\Lambda$CDM models discussed above. The addition of $w$ as a free parameter increases the confidence regions substantially, as it can be seen from the figures and also reported in Tab.~\ref{tab:results}, where the FoM decreases compared to previous cases. Here the variation is due to the parameter itself rather than the addition of an extra parameter; in fact, if we consider the non-flat $\Lambda$CDM model, which has the same number of parameters as $w$CDM model, the FoM reduces from $0.145$ to $0.124$ which corresponds to almost 15\%. However, the 3-FoM manifests an opposite behavior, it increases of about $8\%$. The reason is that $f\sigma_8(z)$ is sensitive to the variation of the parameters almost at any redshift, whereas the Hubble parameter is more sensitive at high redshifts (fixing one value of $H_0$, the variation on $H(z)$ can only appear when the $z$ is increased). For the non-flat $\Lambda$CDM model the area enclosed by $f\sigma_8(z)$ is smaller than the era enclosed for the $w$CDM model, hence giving a lager FoM. However, the maximum value of the Hubble parameter is larger for non-flat $\Lambda$CDM model, $210.06$ against $204.89$ for the $w$CDM model. This effect is taken into account in the 3-FoM, where the errors on $H(z)$ are considered. The two effects are counterbalanced, giving almost the same value in the 3-FoM.

\begin{figure}[H]
 %   \centering
    \includegraphics[width=0.48\textwidth]{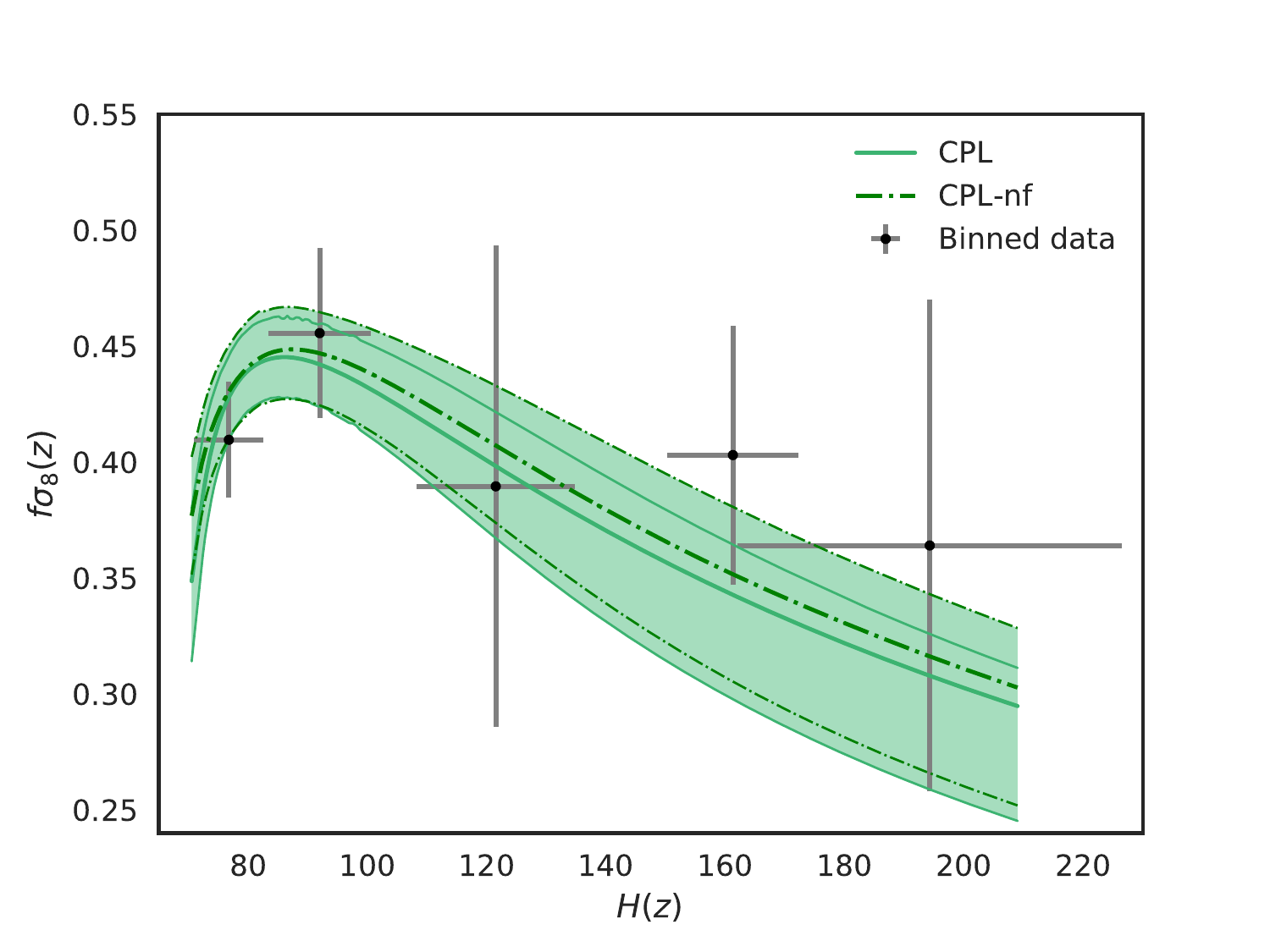}\\
    \includegraphics[width=0.48\textwidth]{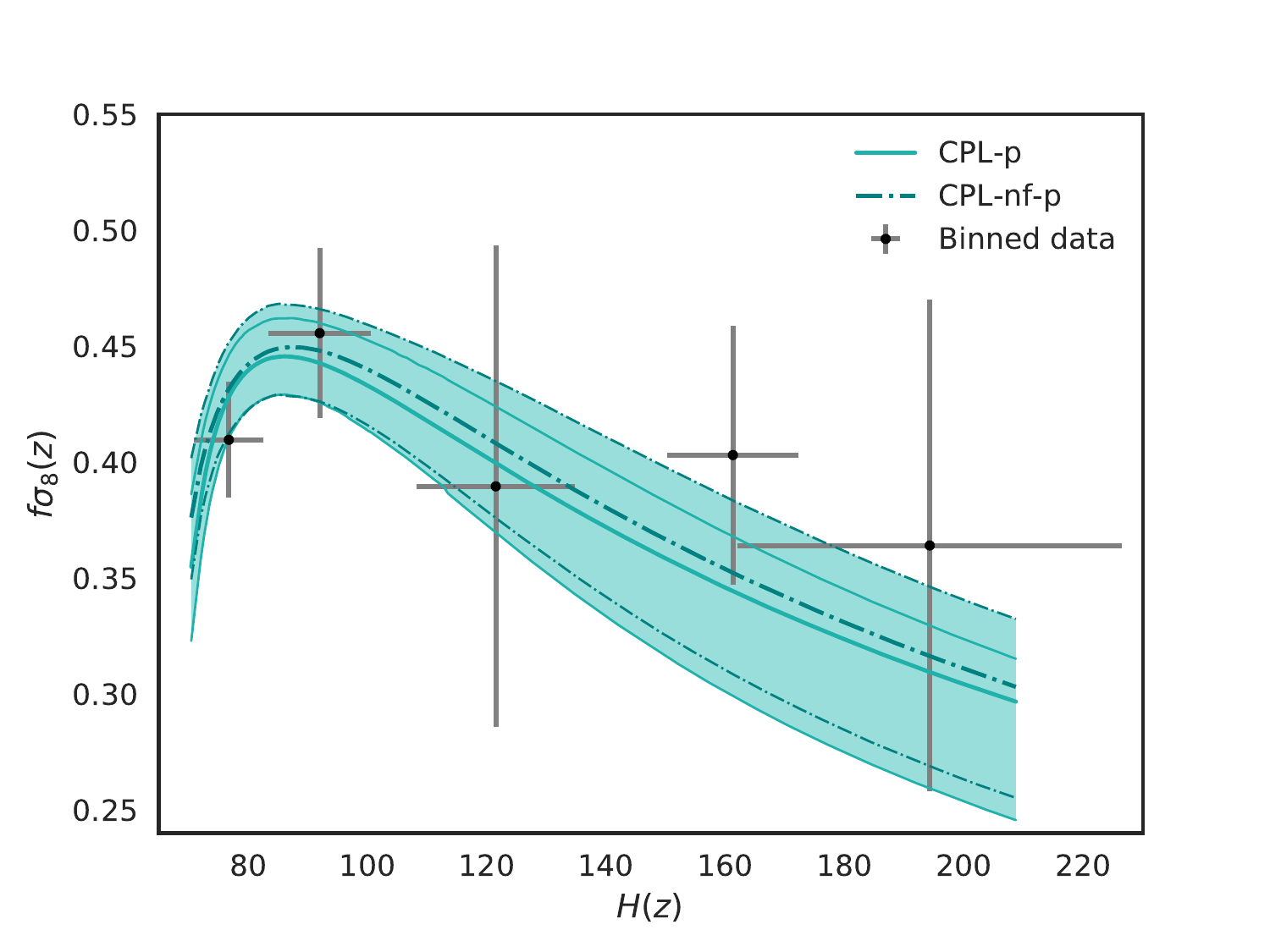}
    \caption{The conjoined plots of the cosmic growth $f\sigma_8(z)$ versus the cosmic expansion $H(z)$ for differents models described in the text: (upper panel) CPL with CPL-nf and (lower panel) CPL-p with CPL-p-nf. Also the 1$\sigma$ error regions (shaded areas) and the real binned data (gray points) are shown.}
    \label{fig:CPL-joint}
\end{figure}

The evidence is weakly in support of the non-flat $\Lambda$CDM model over $w$CDM and the same is found for the BIC and AIC$_c$ criteria. Adding curvature to the $w$CDM model makes the FoM increase of about 1\% meaning that the 1$\sigma$ errors are almost the same, however, their best-fits differ. The 3-FoM decreases for the non-flat model showing that their errors are less constrained. The BIC and AIC$_c$ supports the flat model but the evidence is inconclusive.

\begin{table*}[htp]%[H]%[htp]
\centering
\begin{tabular}{|c|ccccccc|}
\hline
Model & $\Omega_{m_0}$ & $\Omega_{de_0}$ & $w_0$ & $w_a$ & $c_s^2$ & $H_0$ & $\sigma_8$  \\ 
\hline
$\Lambda$CDM & $0.286^{+0.032}_{-0.038}$ & $\bf{1-\Omega_{m_0}}$ & \textbf{-} & \textbf{-} & \textbf{-} & $69.7\pm 2.3$ & $0.779\pm 0.039$ \\
$\Lambda$CDM-nf    & $0.37\pm 0.16$ & $0.83^{+0.29}_{-0.24}$ & \textbf{-} & \textbf{-} & \textbf{-} & $70.4\pm 3.1$ & $0.762^{+0.044}_{-0.084}$\\
$w$CDM & $0.280^{+0.045}_{-0.039}$ & $\bf{1-\Omega_{m_0}}$ & $-1.11^{+0.38}_{-0.30}$ & \textbf{-} & \textbf{-} & $70.6^{+4.2}_{-4.7}$ & $0.782^{+0.045}_{-0.11}$\\
$w$CDM-p & $0.278^{+0.044}_{-0.037}$ & $\bf{1-\Omega_{m_0}}$ & $-1.09^{+0.38}_{-0.30}$ & \textbf{-} & $0.50\pm 0.29$ & $70.5\pm 4.4$ &  $0.788^{+0.045}_{-0.11}$\\
$w$CDM-nf & $0.34^{+0.18}_{-0.22}$ & $0.86^{+0.34}_{-0.41}$ & $-1.08^{+0.49}_{-0.18}$ & \textbf{-} & \textbf{-} & $69.7\pm 4.2$ & $0.790^{+0.045}_{-0.11}$ \\
$w$CDM-nf-p & $0.34^{+0.18}_{-0.22}$ & $0.85^{+0.34}_{-0.42}$ & $-1.07^{+0.50}_{-0.16}$ & \textbf{-} & $0.51\pm 0.29$ & $69.6^{+3.7}_{-4.7}$ & $0.795^{+0.048}_{-0.11}$ \\
CPL & $0.294^{+0.047}_{-0.041}$ & $\bf{1-\Omega_{m_0}}$ & $-1.20\pm 0.34$ & $-0.50^{+0.99}_{-0.46}$ & \textbf{-} & $71.9\pm 4.5$ & $0.747^{+0.026}_{-0.099}$\\ 
CPL-p & $0.293^{+0.046}_{-0.041}$ & $\bf{1-\Omega_{m_0}}$ & $-1.17\pm 0.33$ & $-0.50^{+1.0}_{-0.49}$ & $0.50\pm 0.29$ & $71.7\pm 4.4$ & $0.751^{+0.030}_{-0.10}$ \\ 
CPL-nf & $0.27^{+0.12}_{-0.24}$ & $0.72^{+0.21}_{-0.45}$ & $-1.27^{+0.63}_{-0.27}$ & $-0.42^{+1.0}_{-0.47}$ & \textbf{-} & $70.5^{+4.0}_{-4.6}$ & $0.778^{+0.048}_{-0.11}$ \\
CPL-nf-p & $0.27^{+0.11}_{-0.26}$ & $0.72^{+0.23}_{-0.45}$ & $-1.28^{+0.65}_{-0.27}$ & $-0.38^{+0.98}_{-0.45}$ & $0.50\pm 0.29$ & $70.4^{+3.9}_{-4.5}$ & $0.778^{+0.054}_{-0.10}$\\ 
\hline
\end{tabular}
\caption{Parameter constraints derived from Nested Sampling to each (non-analytical) model described in the text.}
\label{tab:par-constraints}
\end{table*}

In Fig.~\ref{figs:L-w-joint} (lower panel) we show the  reconstruction of the $H(z)-f\sigma_8(z)$ assuming flat and non-flat $w$CDM with the further addition of perturbations in the dark energy sector parameterized with $c_s^2$ as an extra free parameter. If we compare the latest results with the former case we realize that the FoM does not change from $w$CDM to $w$CDM-p, whereas it decreases of about  2.4\% from $w$CDM-nf to $w$CDM-nf-p . These negligible variations are repeated for the 3-FoM that does not change from $w$CDM-nf to $w$CDM-nf-p and it increases of about 7.1\% from flat $w$CDM to $w$CDM-p. As expected, dark energy perturbations are weakly constrained with the data available (dark energy perturbations affect only the growth of matter).  This is shown in Tab.~\ref{tab:par-constraints} where the best fits of the models with and without dark energy perturbations are basically the same. This behavior is shown in all the criteria used in this work, except for the BIC criterion which indeed favors the model without dark energy perturbations. However, this is a pure mathematical effect as the BIC criterion always penalizes the model with extra parameters. 

In Fig.~\ref{fig:CPL-joint} (top panel) are shown the reconstructions of the $H-f\sigma_8(z)$ assuming flat and non-flat CPL. 
If we look at Tab.~\ref{tab:results}, we realized that the FoM constrains better CPL over $w$CDM which might sound peculiar because one would naively expect that a model with more parameters has larger 1$\sigma$ errors. Here, the difference in the FoM comes from the asymmetric values of the errors on $w_a$; this asymmetry is due to the choice of the prior for $w_a$, for which we chose to bind it to $w_0$ in order to guarantee an accelerated expansion.  This asymmetry led to a smaller area in the upper part, reducing the enclosed 1$\sigma$ area of $f\sigma_8(z)$. The 3-FoM is more stable and this is again due to the value of the Hubble parameter at high redshifts: for the $w$CDM model $H(z=2)=204.89$, whereas for CPL model is $208.97$. This $2\%$ difference is accounted in the final 3-FoM which decreases with respect to its companion. The evidence gives inconclusive results, manifesting the negligible effects of $w_a$ on the two observables; the same conclusion is obtained with the AIC$_c$ criterion. However, the BIC criterion strongly penalizes CPL just because of the extra parameter in the model. 

The CPL, $w$CDM-nf, and $w$CDM-nf models have the same number of parameters, thus BIC and AIC$_c$ criteria change less than 1\% between them, showing again that they depend strongly on the number of parameters. The FoM and 3-FoM show that CPL is better constrained but, as mentioned, this is due to the priors on $w_a$ used. The evidence weakly favors the $w$CDM-p model over CPL but it is inconclusive with respect of $w$CDM-nf.

The non-flat CPL have the same number of parameters as $w$CDM-nf-p, but the FoM shows that CPL is better constrained by the data whereas the 3-FoM does not change. The BIC and AIC$_c$ change less than 1\% and the evidence weakly favors $w$CDM-nf-p model.

In Fig.~\ref{fig:CPL-joint} (lower panel) we show the reconstruction of the $H-f\sigma_8(z)$ assuming flat and non-flat CPL-p models. The behavior is similar to the previous case (CPL versus CPL-nf). By adding the curvature parameter the evidence is inconclusive and the other indicators favor the flat model because it has one parameter less.  
When we take into account dark energy perturbations into the CPL models, we obtain a similar behavior as seen for the $w$CDM models. Again, with the available data we are not able to constrain $c_s^2$, hence all the criteria are insensitive to the variation of the sound speed. 
The only exceptions are BIC and AIC$_c$ criteria, which penalize the addition of the sound speed into the analysis. 

For completeness we also performed our analysis using the analytical solutions for the growth rate of matter, the models are $\Lambda$CDM, $w$CDM and $w$CDM with dark energy perturbations. The results are reported in the Appendix \ref{Append} and the results are shown in Tab.~\ref{tab:results-anal}, whereas the best fit of these three models can be found in Tab.~\ref{tab:par-constraints-anal}. 
All the three analytical models give results in excellent agreement with the full numerical analysis, demonstrating that the analytical solutions found in the literature are consistent and they can be safely used. 

\begin{figure} %[H]
   \centering
    \includegraphics[width=0.48\textwidth]{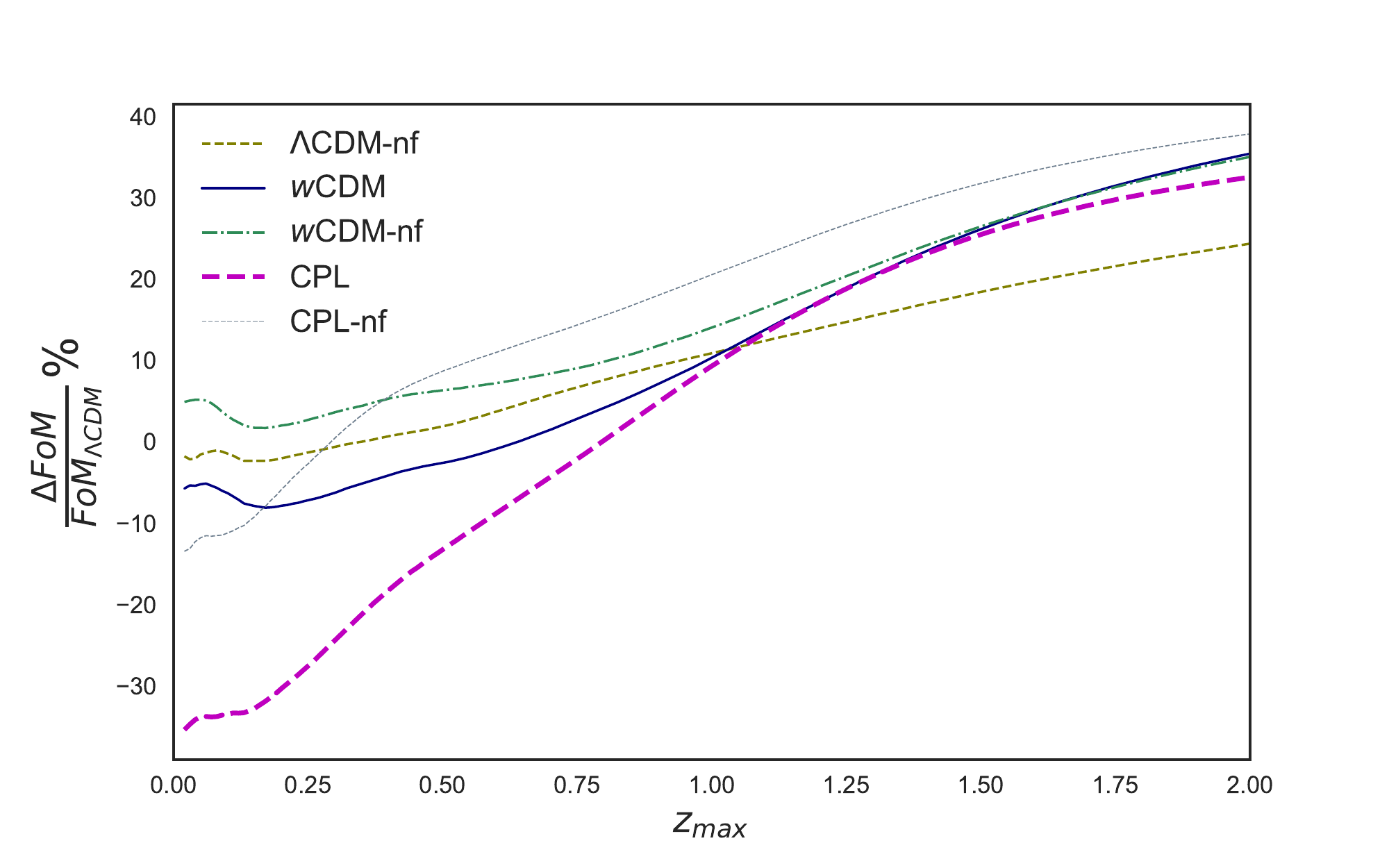}\\
    \includegraphics[width=0.48\textwidth]{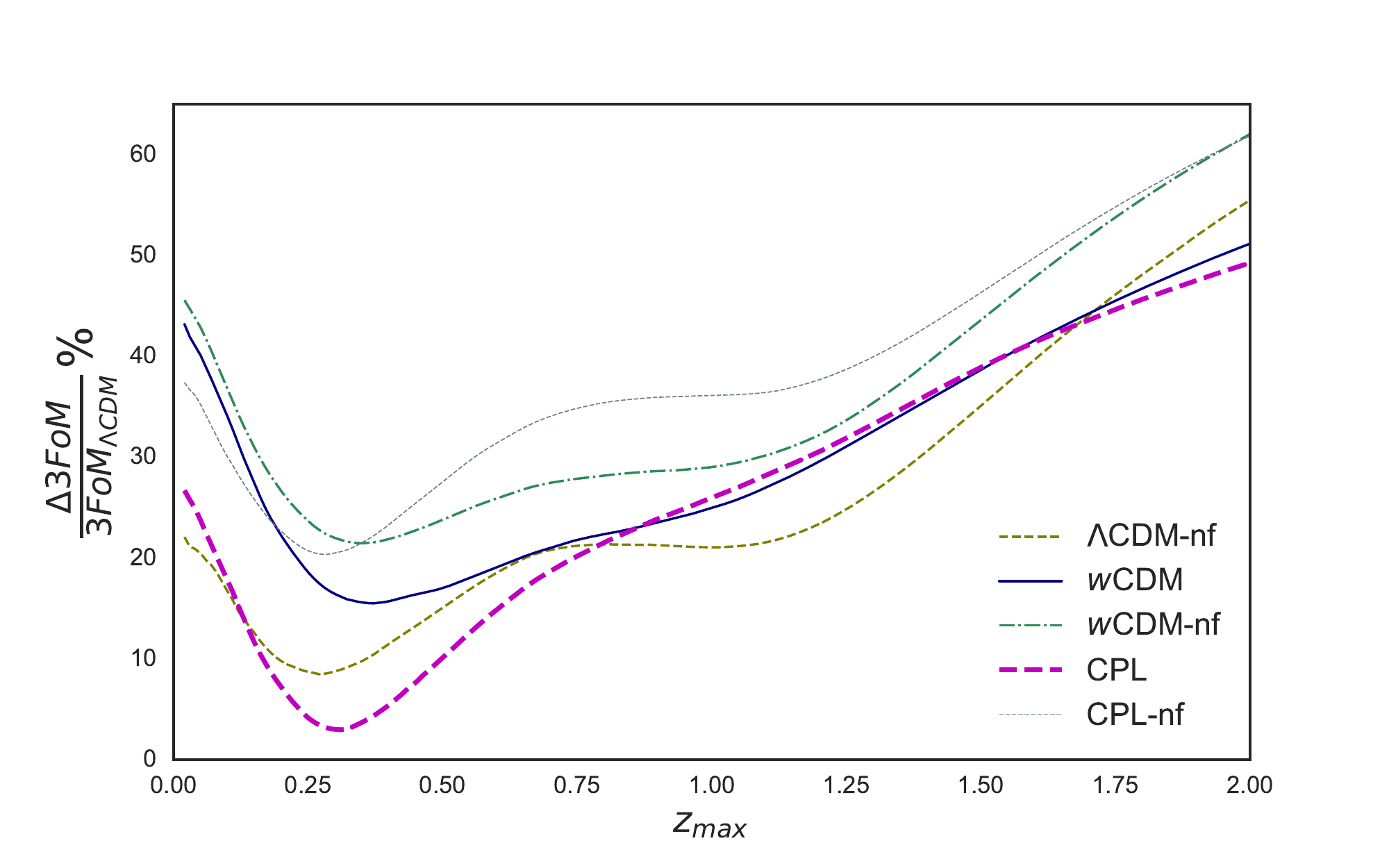}
    \caption{These figures show the percentage difference of FoM (upper panel) or 3-FoM (lower panel) between a model and $\Lambda$CDM. We only present the models without perturbations in the dark sector. Here, $\Delta  \textrm{FoM}=\textrm{FoM}_{\Lambda \textrm{CDM}}-\textrm{FoM}_{\textrm{model}}$ and likewise for the 3-FoM.}
    \label{fig:FoM-3FoM-joint}
\end{figure}

\section{Conclusions}\label{conclusions}

In our work we implemented the conjoined $H(z)-f\sigma_8(z)$ method in order to test an entire family of ten dark energy models; we started with the simplest model, $\Lambda$CDM which is described by three parameters only, and we systematically increased the level of complexity of the model by adding extra parameters, being the non-flat CPL with dark energy perturbation the most complex model (with seven parameters).

For each model, we first found the best fit using MCMC analysis by combining the most recent cosmic chronometer and growth data available.  Subsequently, we compared the dark energy models with five different statistical criteria, aiming at highlighting the potentiality and the weakness of each criterion. 

As expected, we found that the evidence is the most accurate statistical test to compare different models as it takes into account the information of the entire likelihood of the parameters and it does not always penalize a model with extra parameters. The 3-FoM better characterizes the sensitivity of the parameters according to the data used. This criterion takes into account simultaneously the errors from both $f\sigma_8(z)$ and $H(z)$; in particular, we showed that the errors of the Hubble parameter increase with redshift and this has an important effect on the constraining power of the test. The FoM instead is limited only to $f\sigma_8(z)$, hence neglecting the information from $H(z)$, which might be crucial if the analysis is extended at high redshift. As a complementary test, we performed the same analysis in the same redshift range as in~\cite{Basilakos:2017rgc} and we found consistent results. 

For the last two criteria, BIC and AIC$_c$, we showed that they always penalize the addition of extra parameters; in fact, if we consider the two extreme models, i.e. $\Lambda$CDM with only three parameters and non-flat CPL with dark energy perturbations, which has seven parameters, we find that $\Delta \text{BIC} \sim 40$ manifesting a {\em very} strong evidence in favor of the $\Lambda$CDM model. Similarly, but less decisive is $\Delta$AIC$_c$ for which we find a value of $\sim 10$, which still favors strongly $\Lambda$CDM but more moderately than BIC.  
 
To demonstrate the power of the 3-FoM, we compute the FoM and 3-FoM at different redshifts starting from $z=0$ up to the $z_{max}$. 
These results are shown in Fig.~\ref{fig:FoM-3FoM-joint} where we plotted the relative difference of the FoM (top panel) and the 3-FoM (lower panel) for each model with respect to $\Lambda$CDM. It is interesting to notice that at low redshifts the FoM for $w$CDM, $w$CDM-nf, CPL, and CPL-nf is larger than $\Lambda$CDM, meaning that the former is better constrained than the latter. This effect is not manifested in the 3-FoM which is always larger for the $\Lambda$CDM model. 

\begin{table}[h]
\centering
\begin{tabular}{ccc}
\hline
\hline
Redshift bin                    & $H(z)$ [km s$^{-1}$ Mpc$^{-1}$]  & $f\sigma_8(z)$  \\ \hline
0 $< z \leq$ 0.4            & 76.8 $\pm$ 5.8     & 0.410 $\pm$ 0.025 \\
0.4 $< z \leq$ 0.8        & 92.0 $\pm$ 8.6     & 0.456 $\pm$ 0.037 \\
0.8 $< z \leq$ 0.12        & 121.5 $\pm$ 13.3     & 0.390 $\pm$ 0.104 \\
0.12 $< z \leq$ 0.16    & 161.2 $\pm$ 11.0    & 0.404 $\pm$ 0.056 \\
1.16 $< z \leq$ 1.2        & 194.2 $\pm$ 32.2    & 0.364 $\pm$ 0.106 \\
\hline
\hline
\end{tabular}
\caption{Binned measurements of $H(z)$ and $f\sigma_8(z)$ with equispaced redshifts points and its uncertainties. These are the gray points shown in Figs.~\ref{figs:L-w-joint} and \ref{fig:CPL-joint}.} 
\label{tab:binned}
\end{table}

\acknowledgments

BS acknowledges help from CONICYT. JS and DS acknowledge financial support from the Fondecyt project number 11140496. We also thank Savvas Nesseris for useful discussions.

\appendix

\section{Comparison with analytical solutions}\label{Append}

The second order differential equation for the matter density contrast at small scales, without dark energy perturbations is given by, \cite{Belloso:2011ms}
\bea
a^2\delta_m'' +\left(3-\epsilon(a)\right)a\delta_m'-\frac{3}{2}\Omega_m(a)\delta_m(a)=0\;,\nn
\eea
with $\epsilon(a)=-d\log H(a)/d\log a$. As we are describing late time solutions we will always take the growing mode solution given by \cite{Belloso:2011ms}
\be
\delta(a)=a_2F_1\bigg(\frac{w-1}{2w},\,-\frac{1}{3w},\,1-\frac{5}{6w},1-\Omega_m^{-1}(a)\bigg)\,,\nn
\ee
where we omitted the integration constant because it will cancel out when we evaluate $f(a)$. The result to $\Lambda$CDM is given by setting $w=-1$.

There exist analytical solution for the matter density contrast when dark energy perturbations are included, see \cite{Nesseris:2015fqa} for mode details. 
The joint solution for the density contrast is given by
\bea
\delta(a)=a_2F_1\bigg(\frac{1}{4}&-&\frac{5}{12w}+B,\frac{1}{4}-\frac{5}{12w}-B,\nn\\
1&-&\frac{5}{6w},1-\Omega_m^{-1}(a)\bigg)\;,\nn
\eea
where $B$ is used as $B_\text{joint}$ in \cite{Nesseris:2015fqa}, which corresponds to:
\bea
B=\frac{1}{12w}\sqrt{(1-3w)^2+24\frac{1+w}{1-3w+\frac{2}{3}\frac{k^2c_s^2}{H_0^2\Omega_{m_0}}}}\;.\nn
\eea 

\begin{table}[H]%[H]%[h]
\begin{center}
\begin{tabular}{lcccccc}
\hline
\hline
Model &  $\log(E)$ &    FoM &   3FoM & BIC &    AIC$_c$ & $H_{max}$\\
\hline
$\Lambda$CDM-a   & -22.07 &  0.192 & 0.027 & 51.34 & 34.01 & 202.05 \\
$w$CDM-a         & -23.28 &  0.125 & 0.014 & 59.32 & 36.39 & 204.80\\
$w$CDM-p-a       & -23.23 &  0.126 & 0.014 & 67.22 & 38.79 & 204.66\\
\hline
\hline
\end{tabular}
\end{center}
\caption{Results of the different methods for each analytic model. These are almost equal to their numerical versions.}
\label{tab:results-anal}
\end{table}

\begin{table}[H]
\begin{center}
\centering
\begin{tabular}{|c|ccc|}
\hline
 &  $\Lambda$CDM-a  & $w$CDM-a & $w $CDM-p-a \\ 
\hline
$\Omega_{m_0}$ & $0.286^{+0.033}_{-0.038}$ & $0.281^{+0.044}_{-0.039}$ & $0.281\pm 0.044$ \\
$w_0$ & \textbf{-} & $-1.10^{+0.36}_{-0.31}$ & $-1.10^{+0.36}_{-0.30}$ \\
 $c_s^2$ &  \textbf{-} &\textbf{-} &$0.50\pm 0.29$\\
 $H_0$ & $69.8\pm 2.4$ &$70.4\pm 4.4$ &$70.5^{+4.1}_{-4.7}$ \\
 $\sigma_8$ & $0.780\pm 0.040$ & $0.784^{+0.042}_{-0.11}$& $0.784^{+0.044}_{-0.11}$ \\
 \hline
\end{tabular}
\end{center}
\caption{Parameter constraints derived from Nested Sampling to each analytical model described in the text.}
\label{tab:par-constraints-anal}
\end{table}

\bibliographystyle{utphys}
\bibliography{bibliography}

\end{document}